\documentclass[runningheads]{llncs}
\usepackage[dvipsnames,svgnames]{xcolor}

\usepackage{lipsum}

\usepackage{ifxetex}
\ifxetex
  \usepackage{fontspec}
\else
  \usepackage[T1]{fontenc}
\fi

\ifxetex\else\usepackage{cite}\fi

\usepackage{graphicx, placeins, float}

\usepackage{tabularx, booktabs}
\newcolumntype{C}{>{\centering\arraybackslash}X}
\usepackage{arydshln}
\usepackage{hyperref}
\hypersetup{
    colorlinks=true,
    linkcolor=MidnightBlue,
    citecolor=MidnightBlue,
    urlcolor=MidnightBlue
}

\usepackage{amsmath, amssymb, amsfonts, mathtools}

\usepackage{cleveref}
\crefname{ineq}{Inequality}{Inequalities}
\crefname{eq}{Equation}{Equations}

\usepackage{todonotes}
\usepackage{float}
\usepackage{xspace}

\newcommand{\R}{\mathbb{R}}

\newcommand{\beginsupplement}{
        \numberwithin{equation}{section}
        \setcounter{table}{0}
        \renewcommand{\thetable}{S\arabic{table}}%
        \setcounter{figure}{0}
        \renewcommand{\thefigure}{S\arabic{figure}}%
     }

\usepackage{bm}

\def\P{\mathbb{P}}
\def\R{\mathbb{R}}
\def\cN{\mathcal{N}}

\title{A Phylogenetic Approach to Genomic \linebreak Language Modeling}

\author{Carlos Albors\inst{1} \and
Jianan Canal Li\inst{1} \and
Gonzalo Benegas\inst{1} \and
Chengzhong Ye\inst{2} \and
Yun S. Song\inst{1,2}
}

\authorrunning{C. Albors et al.}

\institute{Department of Electrical Engineering and Computer Sciences, University of California, Berkeley, CA 94720, USA  \and 
Department of Statistics, University of California, Berkeley, CA 94720, USA\\
\email{\{calbors, yss\}@berkeley.edu}}

\begin{document}

\maketitle

\begin{abstract}
Genomic language models (gLMs) have shown mostly modest success in identifying evolutionarily constrained elements in mammalian genomes. To address this issue, we introduce a novel framework for training gLMs that explicitly models nucleotide evolution on phylogenetic trees using multispecies whole-genome alignments.  Our approach integrates an alignment into the loss function during training but does not require it for making predictions, thereby enhancing the model's applicability. We applied this framework to train PhyloGPN, a model that excels at predicting functionally disruptive variants from a single sequence alone and demonstrates strong transfer learning capabilities.


\end{abstract}

\section{Introduction}

In recent years, there has been an emerging interest in training large language models on genome sequences \cite{benegas2025genomic}. One of the primary reasons for developing these models is to enable transfer learning. If these models make it possible to interpret genetic variants of otherwise-unknown function, they could make inroads in our understanding of genetics and, in turn, on human health \cite{fowler2023atlas}.

Despite this interest, most genomic language models (gLMs) have generally only achieved modest results at the zero-shot task of identifying deleterious variants in human genomes \cite{benegas2025dna,vishniakov2024genomic}. This is a pivotal task because it reflects the capability of these models to identify functionally important sites. These results are unexpected at first glance because constrained positions can be detected by applying simple statistical methods to the genomes gLMs are trained on \cite{pollard2010phylop,smith2020phylogenetics,yan2023phyloaccgt}. Moreover, they sharply contrast with results for protein language models, which excel at identifying deleterious protein-altering variants \cite{meier2021language,brandes2023genome,jagota2023cross,cheng2023accurate}. These observations suggest that there are unique challenges that standard methods for training gLMs do not fully address.

In recent work, Benegas et al. \cite{benegas2025dna} introduced a gLM named GPN-MSA that achieves cutting-edge performance at this task, even when compared with well-established variant effect predictors. These results are due to a unique aspect of the design of GPN-MSA: it utilizes a multiple sequence alignment (MSA) of diverse vertebrate genomes to the human reference genome as supplementary features, demonstrating that these data are an invaluable resource for modeling human genomes. However,
for making a prediction, it requires as input an MSA for the same set of species as in training data, which makes it difficult to apply the model to other species or to regions where alignment to the human genome is poor.
This undermines its utility for transfer learning.
Furthermore, if genomes that are too similar to the human genome are included in the MSA, the model learns to simply copy from them, resulting in the learned probability distribution not being useful for variant effect prediction.  Therefore, most primates were excluded from GPN-MSA's training data, which is not ideal since they provide useful information about relatively recent evolutionary constraints relevant to humans \cite{gao2023landscape}.   Lastly, GPN-MSA relies on existing conservation annotations \cite{phastcons,pollard2010phylop} to filter training data and re-weight the loss function in order to mitigate the influence of poorly conserved regions on the model. 

To address these drawbacks, we introduce here a novel framework for training gLMs that makes use of whole-genome alignment data during training but does not require them for making predictions. Our basic insight is that language modeling of biological sequences can be viewed as modeling  stationary distributions of site-specific stochastic processes according to which these sequences evolve \cite{weinstein2022nonidentifiability}. In view of this insight, we propose to train a gLM on the task of modeling the evolution of aligned nucleotides given their phylogenetic relationships, and devise a loss function for this purpose. In this way, our framework bridges ideas from classical phylogenetics with contemporary perspectives on how to train context-dependent models that are well-suited for transfer learning \cite{bommasani2021}.

We applied our framework to develop PhyloGPN (Phylogenetics-based  Genomic Pre-trained Network), a gLM that achieves state-of-the-art performance on several transfer learning evaluations, including five out of seven tasks in the BEND set of benchmarks \cite{marin2024bend}. Especially notably, this model performs comparably with state-of-the-art methods for deleteriousness prediction on commonly used benchmarks, vastly outperforming gLMs other than GPN-MSA. In this work, we describe our methods and discuss these results.


\section{Background and Related Work}

\subsection{Transfer Learning and Genomic Language Models (gLMs)}

A practical challenge for model development is that data specific to a task of interest are often too limited to be useful. In this event, it is necessary to regularize models in order to prevent overfitting.
One approach to regularize models is to perform transfer learning---that is, to utilize a model trained to perform an upstream task to support the development of one intended to perform a downstream task. One way to accomplish this is to use some of the intermediate outputs of neural networks at hidden layers as features for other models. These features may be viewed as embeddings of the input data that represent relevant axes of variation. A related technique is ``fine-tuning'': to simply retrain a network that was previously ``pretrained'' on an upstream task to perform the downstream task of interest. By initializing with pretrained weights, fine-tuned models are steered toward parameters that may enable improved generalization performance. These techniques have enabled the development of effective models for specialized tasks \cite{pan2010transfer,mikolov2013word2vec,devlin2018bert}.

An upstream task that may provide useful information for genome interpretation is language modeling on genome sequences: the task of predicting which nucleotides are more likely to occur in the context of a given sequence of nucleotides. Genome sequences are shaped by evolutionary forces---namely, mutation and selection---that determine the frequency with which each nucleotide takes the place of another at each position in a genome sequence. These frequencies, in turn, inform how likely it is that we observe a particular nucleotide at a given position of an extant sequence. As a result, language models trained on these sequences may implicitly learn to distinguish evolutionarily constrained sequences from those that evolve neutrally. Since functionally important elements are necessarily under evolutionary constraint, gLMs can, in principle, be used to support a wide array of genome interpretation tasks.

Various language models of human genomes have been developed with the aim of enabling transfer learning. Most of these use the Transformer architecture \cite{vaswani2017attention} and have been trained with masked language modeling \cite{devlin2018bert} either only on the human reference genome, on thousands of human genomes, or on diverse genomes across the tree of life. This class includes the Nucleotide Transformer \cite{dalla-torre2025nucleotide}, DNABERT \cite{ji2021dnabert,zhou2024dnabert2}, and GENA-LM models \cite{fishman2023genalm}. Other models deviate from this protocol. For instance, the HyenaDNA models use a specialized convolutional architecture that enables them to have an exceptionally large receptive field \cite{nguyen2024hyenadna}. These were trained with causal language modeling on the human reference genome. In the same vein, the Caduceus models \cite{schiff2024caduceus} are bidirectional State Space Models \cite{gu2024mamba} whose architecture also enables a large receptive field and additionally enforces reverse-complement equivariance. They were trained with masked language modeling on the human reference genome.

\subsection{Molecular Phylogenetics}

The basic data that represent evolutionary relationships between aligned sequences are phylogenetic trees. 
The leaves of the tree correspond to extant species that are included in the alignment. The interior nodes, on the other hand, correspond to extinct ancestral species.
In our context,  edges are labeled with values in $[0, \infty)$ that represent the evolutionary distances between nodes. 

The evolution of the nucleotide state at a position is typically modeled with a continuous-time Markov chain, parametrized by a rate matrix. There are several named models of nucleotide substitution that differ in how their rate matrices are constrained and, as a result, by their computational properties \cite{felsenstein2003inferring,yang2006computational}.
For our purposes, a convenient model is the F81 model by Felsenstein \cite{felsenstein1981}. This model is relatively expressive---e.g., it can represent any stationary distribution---and has a simple closed-form expression for its transition probabilities. It is also straightforward to extend it to larger state spaces (e.g., the set of amino acids). Its rate matrix is determined by parameters \mbox{$\lambda_{\mathsf{A}}, \lambda_{\mathsf{C}}, \lambda_{\mathsf{G}}, \lambda_{\mathsf{T}} \in (0, \infty)$}
representing the rates at which each nucleotide replaces any other. Alternatively, it can be parametrized in terms of 
$\theta = (\theta_\mathsf{A},\theta_\mathsf{C},\theta_\mathsf{G},\theta_\mathsf{T})\in \R^4,$
where $\theta_a = \log \lambda_a$ for $a \in \cN = \{\mathsf{A}, \mathsf{C}, \mathsf{G}, \mathsf{T} \}.$ We can derive various quantities of interest from these parameters. For example, the stationary probability of the state $a \in \cN$ is given by
$\pi_a = \lambda_a/\sum_{b \in \cN} \lambda_{b} = \operatorname{softmax}(\theta)_a,$
and the probability of at least one substitution occurring in an interval of time 
$t$ is given by
\begin{equation}
\label{eq:probability_of_substitution}
\alpha(t) = 1 - \exp \bigg(-t \sum_{a \in \cN} \lambda_a \bigg) = 1 - \exp \bigg(-t \sum_{a \in \cN} e^{\theta_a} \bigg).
\end{equation}
The transition probabilities of the model are given by
\begin{equation}
\label{eq:f81_transition_probability}
\P_{\mathsf{F81}}(a \mid \theta, b, t) =
\begin{cases}
\alpha(t) (\pi_a - 1) + 1 & \text{if} \ a = b, \\
\alpha(t) \pi_a & \text{otherwise}.
\end{cases}
\end{equation}

The F81 model can be used to derive a likelihood function for the nucleotides at a set of aligned sites given a phylogenetic tree relating these nucleotides. First, it is necessary to specify the distribution of the state corresponding to the root. A convenient choice is to regard this state as a sample from the stationary distribution. The likelihood of the observed nucleotides can then be obtained by using the sum-product algorithm---known as Felsenstein's pruning algorithm in this context \cite{felsenstein1973maximum}---to marginalize out the latent states. This procedure has time complexity on the order of the number of leaves in the tree.



\section{Methods}


\subsection{Training Data}
\label{subsec:training_data}

We compiled a dataset $\mathcal{D}$ where the $i$-th item consists of a possibly padded length-481 substring $x^{(i)}$ of the human reference genome GRCh38,  centered at the $i$-th position of the genome, a list $y^{(i)}$ of nucleotides from different species that are aligned to the $i$-th position, and a phylogenetic tree $T^{(i)}$ whose leaves correspond to the elements of $y^{(i)}$.
For each position $i$, we obtained $y^{(i)}$ from a whole-genome alignment of 447 placental mammalian genomes that was generated by the Zoonomia Consortium by combining an existing alignment of 241 genomes \cite{zoonomia2020} with an additional set of primate genomes \cite{kuderna2024}. Since we require at most one nucleotide from a species for each position in the reference genome, we removed sequences from duplicated species in each alignment block, keeping the sequences that had the smallest edit distance to the consensus sequence of the block and breaking ties by choosing the first sequence to appear in the file.


To obtain $T^{(i)}$, we first downloaded a species-level tree that includes each species in our whole-genome alignment and has been previously used to derive constraint scores from the Zoonomia Consortium's alignment \cite{sullivan2023leveraging}. We then set $T^{(i)}$ to be the minimum spanning tree containing the nodes corresponding to the species that have alignments to this position.

\subsection{Loss Function}

\begin{figure}[t]
\centering
\includegraphics[width=0.7\textwidth]{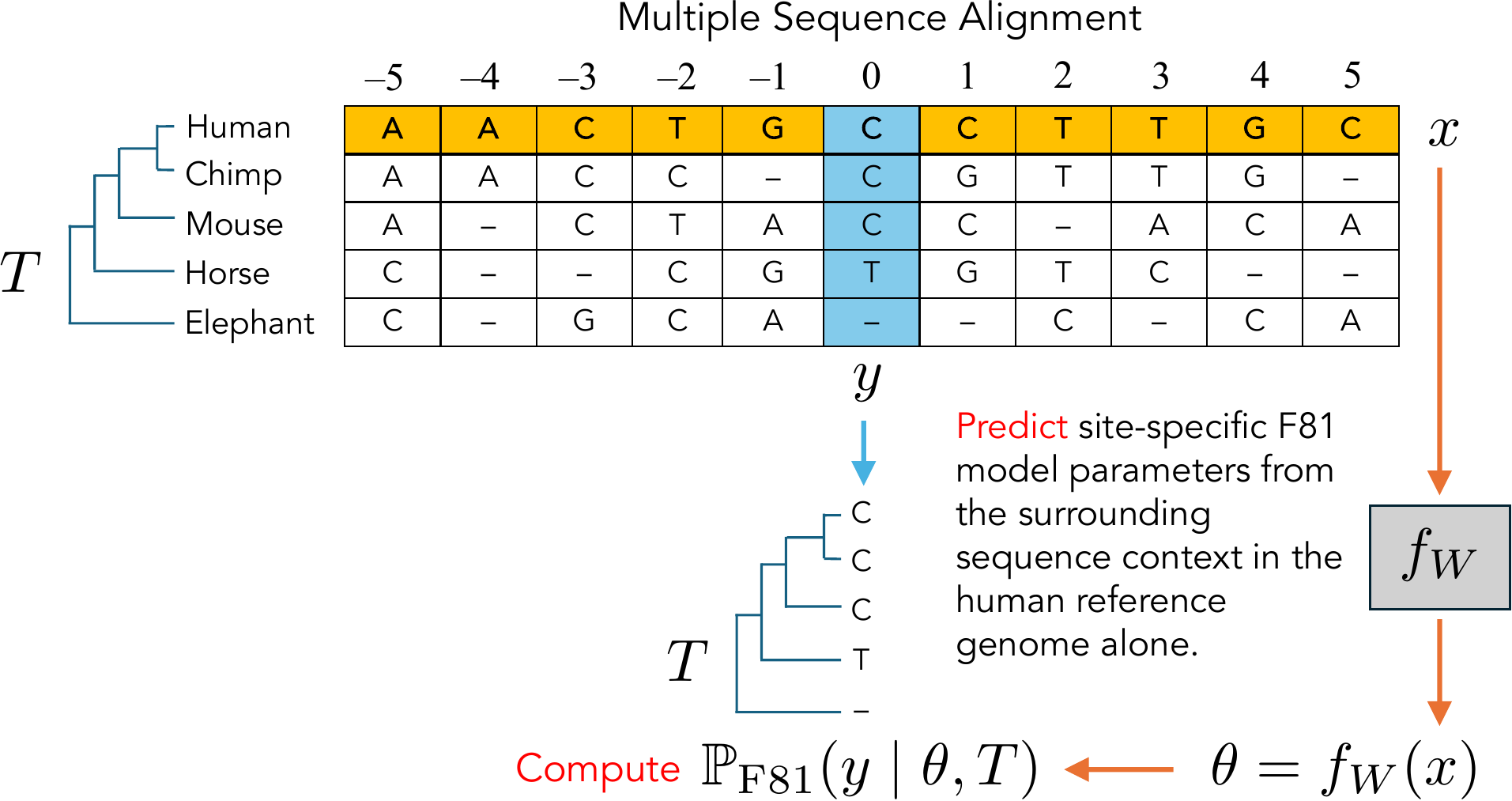}
\caption{Illustration of PhyloGPN's modeling framework. The input data consist of 481~bp windows from the human reference genome GRCh38 and the alignment columns are obtained from a whole-genome alignment of 447 mammalian species to GRCh38.}
\label{fig:framework}
\end{figure}

Formally, PhyloGPN is a neural network $f_W$ with weights $W$ that takes a DNA sequence $x^{(i)}$ of length 481 as input and outputs the parameters of an F81 model for the central position of this sequence (i.e., the $i$-th position of the genome). A preliminary way to train this network is to minimize the loss function
\begin{equation}
\label{eq:loss_func0}
\mathcal{L}_{0}(W; \mathcal{D}) = -\frac{1}{n} \sum_{i = 1}^n \log \P_{\mathsf{F81}} \left( y^{(i)} \;\middle|\; f_W(x^{(i)}), T^{(i)} \right)
\end{equation}
with respect to $W$, as is illustrated in \autoref{fig:framework}.

One issue with applying Eq.~\eqref{eq:loss_func0} to our data is that the central nucleotide of $x^{(i)}$ is also included in $y^{(i)}$. Consequently, models trained to minimize this function may overestimate the importance of the central nucleotide as a feature. To resolve this issue, we conditioned the likelihood 
on the state of the reference nucleotide at the $i$-th position. Letting $\pi^{(i)}(f_W)$ be the probability of the reference nucleotide at this position under the stationary distribution of the F81 model with parameters $f_W(x^{(i)})$, 
conditioning yields the following loss function:
\begin{equation}
\label{eq:loss_func}
\mathcal{L}(W; \mathcal{D}) = \mathcal{L}_{0}(W; \mathcal{D}) + \frac{1}{n} \sum_{i=1}^n \log \pi^{(i)}(f_W).
\end{equation}

Another issue is that Eq.~\eqref{eq:loss_func} is numerically unstable to optimize. The cause of this is the double exponential term in Eq.~\eqref{eq:probability_of_substitution}, which is susceptible to numerical overflow. Our solution is to instead minimize a stable upper bound of Eq.~\eqref{eq:loss_func}. To obtain this bound, we observe that the probability of a transition occurring in a time interval $t$ satisfies the lower bound
$
\alpha(t) \geq \operatorname{sigmoid}\left( \log t + \sum_{a \in \cN} \theta_a\right),
$
which follows immediately from \autoref{prop:sigmoid_bound} in Appendix~\ref{app:math}. In turn, we have 
\begin{equation}
\label{eq:f81_lower_bound}
\P_{\mathsf{F81}}(a \mid \theta, b, t) \geq
\operatorname{sigmoid}\left( \log t + \sum_{a \in \cN} \theta_a \right) \pi_a \quad \text{for} \ a \neq b,
\end{equation}
which determines a lower bound for off-diagonal transition probabilities in Eq.~\eqref{eq:loss_func0} and, therefore, an upper bound for the corresponding terms in Eq.~\eqref{eq:loss_func}. We trained PhyloGPN to minimize this bound.

\subsection{Architecture and Training}

\begin{figure}[t]
\centering
\includegraphics[width=0.88\textwidth]{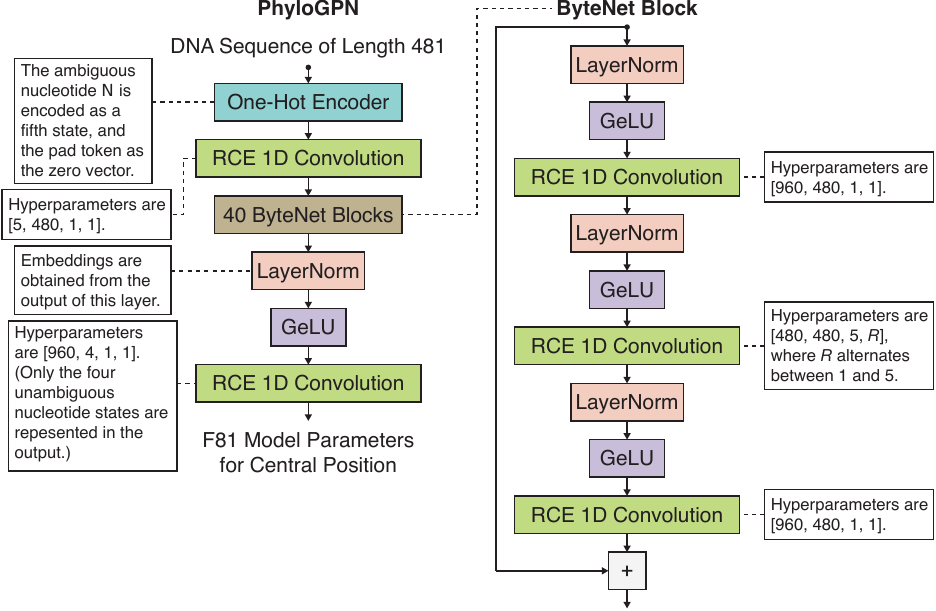}
\caption{Architecture of PhyloGPN. A convolutional layer with $N$ input channels, $M$ output channels, a kernel size $K$, and a dilation rate $R$ is noted as having hyperparameters $[N, M, K, R]$. The acronym ``RCE'' stands for ``reverse-complement equivariant.''
} \label{fig:architecture}
\end{figure}

The architecture of PhyloGPN was adapted from that of CARP \cite{yang2024}---a convolutional neural network (CNN) used for protein language modeling that is itself adapted from ByteNet \cite{kalchbrenner2016neural}. There are two salient differences between these models. First, we do not apply padding to the inputs of the convolutional layers (such that each layer reduces the length of its input by the kernel size minus 1). Second, we enforce reverse-complement equivariance (RCE) by constraining the weights of the network \cite{shrikumar2017}.
PhyloGPN has 40 residual blocks that apply dilation to expand its receptive field to 481 bp. In total, it has about 83~M parameters.
Due to our RCE parametrization, the number of free parameters is half the number of weights and biases, plus the number of layer normalization parameters. \autoref{fig:architecture} describes the architecture of PhyloGPN in greater detail.

To train PhyloGPN, we merged sets of contiguous blocks in the Zoonomia Consortium's whole-genome alignment into non-overlapping MSAs of approximately 10~kb in length. We sampled 12 MSAs per batch. During each epoch, we sample autosomal positions four times, positions on the X chromosome three times, and positions on the Y chromosome once, so that male and female genotypes are weighed equally. We did not include positions outside the primary assembly in the training data.

We trained PhyloGPN on four NVIDIA A100 GPUs for a total of eighteen epochs using the AdamW optimizer. We used a fixed learning rate of $1 \times 10^{-5}$ and did not apply weight decay. For all other optimization parameters, we used the default values included in PyTorch.

\subsection{Embeddings}

PhyloGPN can be used to generate a 960-dimensional embedding of any 481~bp DNA sequence. These embeddings are obtained from the layer noted in \autoref{fig:architecture}. We did not experiment with embeddings from different layers.

In certain cases where the relatively small receptive field of PhyloGPN was limiting, we found it beneficial to expand its effective receptive field by pooling embeddings of contiguous positions. For a given integer $\ell \geq 0$, let $\varphi_{-\ell}, \dotsc , \varphi_{\ell} \in \R^d$ be a set of $d$-dimensional embeddings for the $2\ell + 1$ positions flanking and including a given position in the center, zero padding whenever necessary, and let $A$ be a fixed $d\times d$ random matrix with entries that are i.i.d. as $\mathcal{N}(0, 1).$ We found it effective to apply the transformation 
\begin{equation}
\label{eq:embedding_transformation}
\varphi_0 \mapsto \varphi_0 + \frac{1}{2\ell + 1} \sum_{i = -\ell}^\ell A \varphi_i,
\end{equation}
where the application of $A$ ensures that the transformed embedding is parallel to $\varphi_0$ in expectation. To apply Eq.~\eqref{eq:embedding_transformation} to the embeddings of PhyloGPN, we set $\ell$ equal to 2,760, such that each transformed embedding pools embeddings from 5,521 positions and, therefore, has an effective receptive field of 6,001~bp. We refer to the transformed embedding function as ``PhyloGPN-X'' (where ``X'' stands for ``expanded'').

\section{Results}
\label{sec:results}

\subsection{Log Likelihood Ratio (LLR) Evaluation}

Our first set of benchmarks assesses the extent to which gLMs can identify deleterious substitutions from the likelihood values they output to perform their training tasks.  Since deleterious alleles are less likely to occur in viable genomes, their likelihood values should be relatively lower than those of neutral alleles.

For these evaluations, we benchmarked PhyloGPN against Nucleotide Transformer, HyenaDNA, and Caduceus. We describe how we evaluated these models in Appendix~\ref{app:vep_evaluation_details}.
We could not benchmark the DNABERT models because the weights for the final layers of these models are not available. We also could not benchmark GENA-LM because its usage of byte-pair encoding \cite{sennrich2016neural} makes it incompatible with our evaluation framework.

For each model, we computed likelihood values for each position in the human reference genome by evaluating the model on the context around these positions. Afterward, for each substitution, we obtained the LLR of the alternate allele over the reference. To evaluate PhyloGPN, we used the probability of each state under the stationary distribution of the predicted nucleotide substitution model as the likelihood value for the state. We describe how we obtained likelihood values for other models in Appendix~\ref{app:vep_evaluation_details}.

\begin{figure}[t]
\centering
\includegraphics[width=0.8\textwidth]{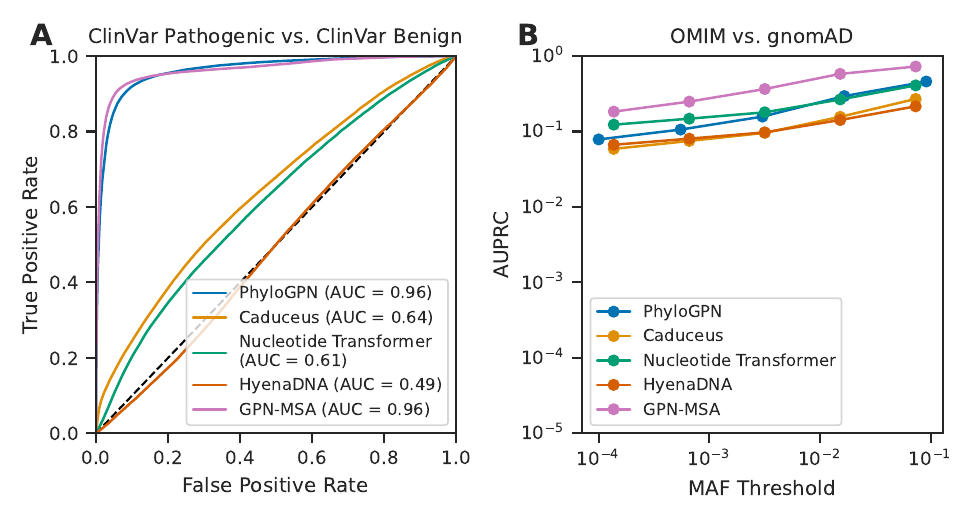}
\caption{Results on predicting clinical labels from LLRs. (a) Results on ClinVar. (b) Results on classifying pathogenic regulatory variants. We show AUPRC results for various negative sets of SNPs above a minimum MAF threshold.}
\label{fig:clinical_results}
\end{figure}

\begin{table}[t]
\centering
\caption{LLR-based results on predicting ClinVar classifications for each variant class. Each entry is an AUROC. The best result for each class is typeset in boldface.}
\label{tab:clinvar_results}
{\scriptsize
\begin{tabularx}{\textwidth}{lCCCC!{\vrule width 0.5pt}C}\toprule
 &  &  & Nucleotide & & \\
Class & PhyloGPN & Caduceus & Transformer & HyenaDNA & GPN-MSA \\
\midrule
3' UTR & \textbf{0.95} & 0.63 & 0.66 & 0.48 & \textbf{0.96} \\
5' UTR & \textbf{0.95} & 0.61 & 0.55 & 0.48 & \textbf{0.95} \\
Downstream of Transcript & \textbf{0.99} & 0.72 & 0.58 & 0.55 & \textbf{0.99} \\
Intronic & \textbf{0.95} & 0.64 & 0.61 & 0.48 & 0.94 \\
Missense & \textbf{0.85} & 0.55 & 0.64 & 0.54 & \textbf{0.91} \\
Noncoding & \textbf{0.95} & 0.59 & 0.54 & 0.47 & \textbf{0.95} \\
Splice Acceptor & \textbf{0.82} & 0.61 & 0.62 & 0.51 & 0.81 \\
Splice Donor & \textbf{0.82} & 0.68 & 0.67 & 0.52 & \textbf{0.83} \\
Start Codon & \textbf{0.80} & 0.41 & 0.69 & 0.43 & \textbf{0.80}  \\
Stop Lost & \textbf{0.82} & 0.43 & 0.55 & 0.48 & \textbf{0.86} \\
Synonymous & \textbf{0.86}  & 0.55 & 0.55 & 0.43 &  \textbf{0.87} \\
Upstream of Transcript & \textbf{0.75} & 0.40 & 0.31 & 0.51 & \textbf{0.84} \\ \bottomrule
\end{tabularx}
}
\end{table}

\smallskip\noindent
\textbf{Classifying ClinVar variants.} 
We first evaluated the extent to which LLRs from each model can distinguish two classes of variants when one of these classes is enriched for deleterious variants. Our first test set was sourced from clinically classified variants in  ClinVar \cite{landrum2017clinvar}. We included all variants that received at least one star and are labeled as pathogenic or likely pathogenic in our positive class, and those that are labeled benign or likely benign in our negative class.
In \mbox{\autoref{fig:clinical_results}a}, we show ROC curves for each model, and in \autoref{tab:clinvar_results}, we report the performance of each model on this task for various subsets of these test data. Our results show that PhyloGPN substantially outperforms our baselines for every category of variants. It also performs comparably to GPN-MSA.

\begin{figure}[t]
\centering
\includegraphics[width=\textwidth]{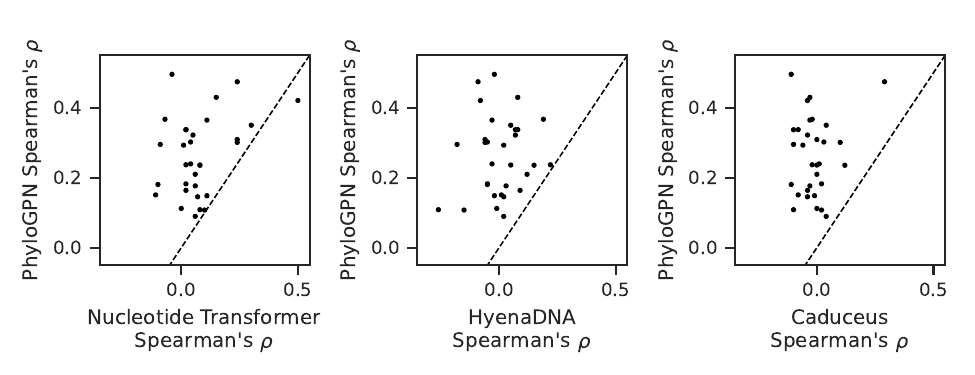}
\caption{Comparison of results for PhyloGPN and baseline models on the task of ranking substitutions in DMS experiments. Each point corresponds to an experiment. The dotted lines are $y = x$ lines.} 
\label{fig:dms_results}
\end{figure}

\smallskip\noindent
\textbf{Discriminating pathogenic regulatory variants from common variants.}
In our second test set, our positive class consists of 409 disease-causing variants sourced from the Online Mendelian Inheritance in Man (OMIM) \cite{smedley2016whole}
database and processed by Benegas et al. \cite{benegas2025dna}. We compared this set with multiple negative sets, each obtained by restricting the set of variants in the gnomAD database \cite{chen2024gnomad} to those above various minor allele frequency (MAF) thresholds. For this test set, the variants in the negative class are encoded with the major allele as the reference and the minor as the alternate. In comparison to ClinVar, the positive class in OMIM is enriched for variants in non-coding sites, and the negative set is less susceptible to ascertainment bias. 
We use AUPRC as our evaluation metric because of the extreme class imbalance in our test set. Our results on this task are shown in \autoref{fig:clinical_results}b. At every MAF threshold, PhyloGPN outperforms all baselines. However, GPN-MSA performs best overall by a very wide margin. 

\smallskip\noindent
\textbf{Predicting deep mutational scanning (DMS) outcomes.}
We also examined how well LLRs from gLMs can rank the results of DMS assays. These assays evaluate the effects of substitutions on the function of a protein; their results indicate whether substitutions are deleterious. Our data were obtained from various published studies and compiled by  Schubach et al. \cite{schubach2024cadd}. Our results are summarized in \autoref{fig:dms_results} and fully described in \autoref{tab:dms_results}. For 24 out of 25 proteins, PhyloGPN outperforms all baseline models, mostly by a wide margin.


\subsection{Embedding Evaluation}

For our second evaluation, we sourced a set of benchmarks from BEND \cite{marin2024bend}, a recent benchmarking suite consisting of seven standard tasks in genomics for evaluating the capability of gLM embeddings to classify elements in human genomes. Five tasks---Gene Finding, Enhancer Annotation, Chromatin Accessibility, Histone Modification, and CpG Methylation---entail training a two-layer CNN on the embeddings of input sequences and evaluating its predictions on test sets. The sixth task---Expression Variant Effect Prediction (VEP)---uses the cosine distance between pairs of embeddings of sequences that differ by a SNP to predict if these SNPs are associated with differential gene expression. Similarly, the final task---Disease VEP---uses this score to predict if SNPs are considered to be pathogenic.

In \autoref{tab:bend_results}, we reproduce results included in Marin et al. \cite{marin2024bend} and include the results of our own evaluation of PhyloGPN, PhyloGPN-X, and the largest RCE Caduceus model. Since the input sequences in BEND are sourced from the human genome, we were also able to include GPN-MSA as a baseline for most of the tasks (see Appendix~\ref{subsec:gpn-msa-bend} for details on our evaluation). We also reproduce results for Expert Methods reported in Marin et al.; these are models that are specifically designed to perform one of the tasks.

Our results show that PhyloGPN has relatively powerful and broad transfer learning capabilities. Most strikingly, on the Disease VEP task, PhyloGPN obtains an AUROC of 0.98. This is an improvement of 0.21 over the next-best model (Nucleotide Transformer) and of 0.43 over the median. In addition, PhyloGPN achieves state-of-the-art performance on Chromatin Accessibility, Histone Modification, and CpG Methylation tasks, matching or surpassing both gLM baselines and Expert Methods.

\begin{table}[t]
\centering
\caption{Embedding-based results on BEND tasks. We report the best result among models in each family. For each task, the best result among gLMs other than GPN-MSA is typeset in boldface. If GPN-MSA or an Expert Method perform best for a task, their results are also in boldface.  On VEP tasks, we also include results on GPN-MSA LLRs between reference and alternative alleles. Asterisked results are discussed further in the text; we believe they are not fully indicative of the models' capabilities. }
\label{tab:bend_results}
{\scriptsize
\begin{tabularx}{\textwidth}{lCCCCCCC}
    \toprule
    \; & Gene Finding & Enhancer Annotation & Chromatin Accessibility & Histone {Modification} & CpG Methylation & Expression VEP & Disease VEP \\
    Model & (MCC) & (AUPRC) & (AUROC) & (AUROC) & (AUROC) & (AUROC) & (AUROC) \\
    \midrule
    PhyloGPN & \phantom{*}0.43* & 0.04 & \textbf{0.86}& \textbf{0.81} & \textbf{0.95} & \phantom{*}0.46* & \textbf{0.98} \\
    PhyloGPN-X & \textbf{0.69} & \phantom{*}0.01* & 0.81 & 0.79 & 0.94 & 0.46 & \textbf{0.98}  \\
    Caduceus & 0.54 & 0.03 & 0.79 & 0.78 & 0.91 & 0.53 & 0.54 \\
    DNABERT & 0.43 & 0.03 & 0.85 & 0.79 & 0.91 & \textbf{0.60} & 0.56 \\
    GENA-LM & 0.52 & 0.04 & 0.84 & 0.78 & 0.91 & 0.49 & 0.55 \\
    HyenaDNA & 0.35 & 0.03 & 0.84 &  0.76 & 0.91 & 0.51 & 0.45 \\
    Nucleotide Transformer\hspace{-3mm} & 0.68 & \textbf{0.06} & 0.80 & 0.78 & 0.92 & 0.54 & 0.77 \\
    \\[-1em]
    \hdashline
    \\[-1em]
    GPN-MSA & 0.38 & 0.03 & N/A & 0.80 & \textbf{0.96} & 0.44 & 0.33 \\
    GPN-MSA (LLR) & N/A & N/A & N/A & N/A & N/A & 0.56 & \textbf{0.98} \\
    Expert Method & \textbf{0.80} & \textbf{0.07} & 0.85 & 0.74 & 0.93 & \textbf{0.70} & 0.56 \\
    & (Augustus) & (Enformer) & (Basset) & (Basset) & (Basset) & (DeepSEA) & (DeepSEA) \\
    \bottomrule
\end{tabularx}
}
\end{table}

\begin{figure}[t]
\centering
\includegraphics[width=\textwidth]{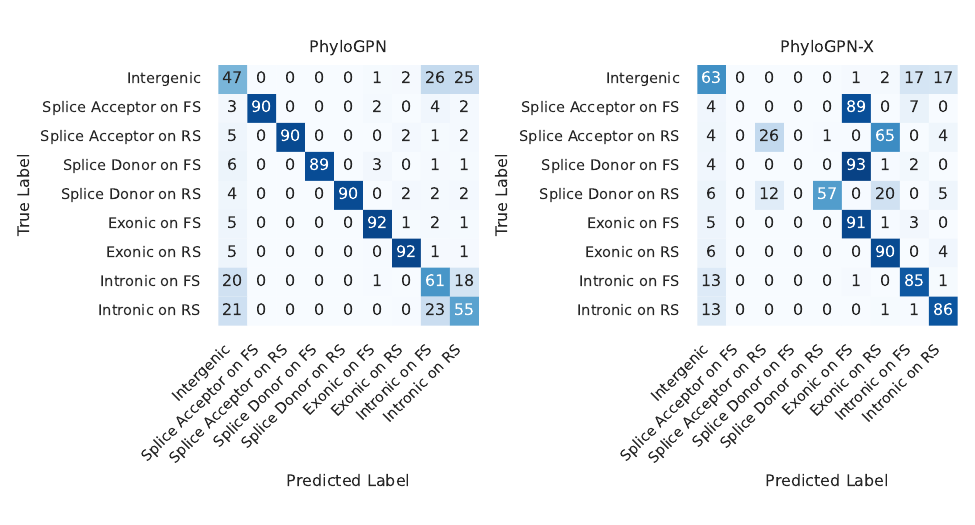}
\caption{Confusion matrices for the Gene Finding task for PhyloGPN and PhyloGPN-X. Entries in cells are the percentage of instances with a true label that were predicted to have a given label. The acronyms ``FS'' and ``RS'' stand for ``forward strand'' and ``reverse stand,'' respectively.}
\label{fig:gene_finding_confusion_matrix}
\end{figure}

However, PhyloGPN does not achieve state-of-the-art performance on certain tasks in BEND. On the Gene Finding task, for instance, PhyloGPN performs below the median. However, on this task PhyloGPN-X surpasses Nucleotide Transformer, the previously best-performing gLM. The confusion matrices shown in \autoref{fig:gene_finding_confusion_matrix} suggest that a larger context size helps PhyloGPN-X discriminate between intronic sites on the forward and reverse strands. Since these sites are easily distinguished based on their proximity to splice sites, we surmise that having a larger context size makes this information available to the classifier. On the other hand, based on \autoref{fig:gene_finding_confusion_matrix}, PhyloGPN-X entirely fails to identify splice donor and splice acceptor sites on the forward strand, misclassifying them as exonic sites. PhyloGPN, in contrast, identifies these sites with approximately 90\% accuracy. We suspect that this discrepancy is due either to the classifier used by BEND being underpowered or the training set being too limited, rather than because the embeddings provide any less information. Indeed, out of 35,749,767 positions in the training data, only 0.11\% are splice sites, whereas 12.45\% are in exons, 48.34\% are in introns, and 39.10\% are in intergenic regions.

On the Enhancer Annotation task, on the other hand, PhyloGPN-X performs notably worse than PhyloGPN. This is the only supervised task for which the classifier used by BEND has a hidden size of 2 instead of 64. Since PhyloGPN-X aggregates a large set of embeddings, we suspected that a more powerful classifier was needed to make use of its outputs. Indeed, as we trained PhyloGPN-X on this task, its loss was relatively static compared to that of PhyloGPN, confirming that the classifier was underfitting the training data. To get a better sense of the representational power of embeddings obtained from PhyloGPN-X, we expanded the hidden size of the classifier to 32. In this case, PhyloGPN-X obtains an AUPRC of 0.03.

Finally, on the Expression VEP task, PhyloGPN performs poorly---a seemingly perplexing result given its strong performance on the Disease VEP task. We hypothesized that PhyloGPN has learned to generate similar embeddings for elements that differ by an eQTL because these elements tend to have similar fitness effects. For example, two regulatory elements that bind to the same set of transcription factors may be  evolutionary constrained in similar ways despite having different binding affinities. On the other hand, elements that differ by a loss-of-function variant---e.g., a gene and its paralogous pseudogene---are constrained very differently because nonfunctional elements evolve neutrally. To test this hypothesis, we evaluated how well the LLRs generated by PhyloGPN could classify variants in the Expression VEP dataset. On this evaluation, PhyloGPN achieved an AUROC of 0.53. By comparison, Nucleotide Transformer---the best-performing model on the Expression VEP task that we could evaluate in a comparable way---achieved an AUROC of 0.36 on this evaluation. These results imply that, despite its result on the Expression VEP task, PhyloGPN may be in some respects more suitable to perform certain downstream tasks that pertain to predicting variant effects on gene regulation.

On four tasks, GPN-MSA shows relatively worse performance than PhyloGPN. The results on the Gene Finding task are likely caused by GPN-MSA having a relatively tiny 128~bp receptive field. On the other hand, the results on the VEP tasks show that even though GPN-MSA's LLRs capably predict the labels of variants in BEND's test sets, the cosine distances between its embeddings of reference and alternate sequences do not.
However, recent work has shown that valuable information can still be extracted from GPN-MSA embedding distances across specific latent dimensions \cite{traitgym}.

\begin{figure}[b!]
\centering
\includegraphics[width=\textwidth]{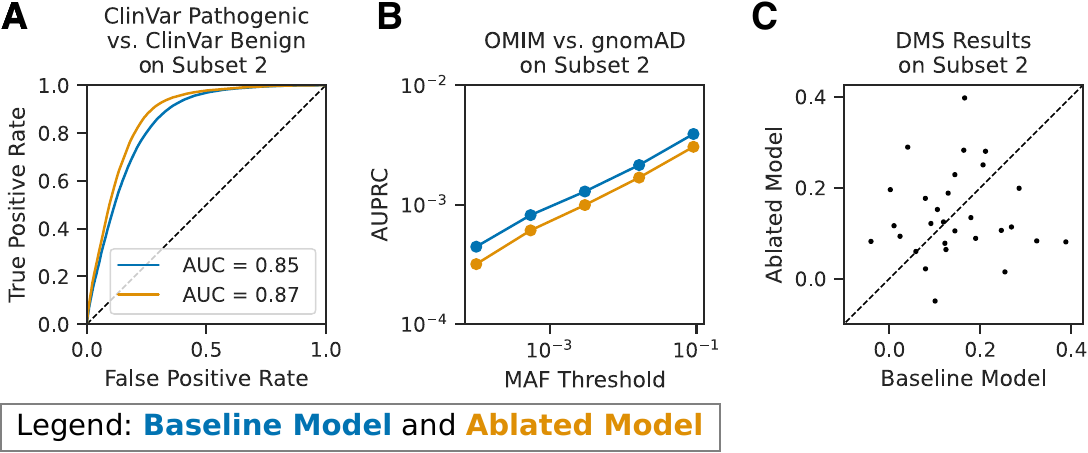}
\caption{Results of ablation study. Subset 1 consists of all positions in the training data of PhyloGPN that are in odd-numbered chromosomes or the X chromosome. Subset 2 consists of all other positions in PhyloGPN's training data. The baseline model is the epoch-one checkpoint of PhyloGPN. The ablated model was trained from scratch for two epochs on Subset 1. (a) Results on classifying ClinVar labels. (b) Results on classifying pathogenic regulatory variants. (c) Results on predicting DMS outcomes.}
\label{fig:ablation_results}
\end{figure}

\subsection{Ablation Study of Generalization}

To determine how well PhyloGPN can generalize to unseen genomic sequences, we conducted an ablation study in which we evaluated a version of PhyloGPN on held-out sequences from the human genome. To do this, we first split the training data of PhyloGPN into two subsets. The first subset, which we denote by Subset 1, consists of all positions in odd-numbered chromosomes and the X chromosome. The second subset, which we denote by Subset 2, consists of all positions in even-numbered chromosomes and the Y chromosome. We then trained a model from scratch on Subset 1 for two epochs and benchmarked it on Subset 2 using our LLR-based evaluations against the epoch-one checkpoint of PhyloGPN. This experimental design ensures that both models see approximately the same amount of training data. The results of this study---presented in \autoref{fig:ablation_results} and \autoref{tab:ablation_results}---show that the ablated model performs comparably to the epoch-one checkpoint. This suggests that PhyloGPN bases its predictions on broadly relevant features of sequences.

\section{Discussion and Conclusion}

In this work, we proposed to train a gLM to model nucleotide evolution. We derived a loss function for this purpose that makes direct use of phylogenetic data. We applied this loss function to train PhyloGPN, a model that is the first of its kind to predict functionally disruptive variants  with high accuracy from a single sequence alone.

One advantage of our framework is that it provides a principled way to control for phylogenetic correlations between genomes in training data \cite{felsenstein2003inferring,yang2006computational}. Indeed, orthologous sites in these genomes are likely to share their state because they inherited it from a common ancestor, and not necessarily because it is functionally important. To mitigate the effects of this phylogenetic bias, previous gLMs---including Nucleotide Transformer and GPN-MSA---have avoided training on genomes from closely related species, discarding an invaluable source of information \cite{kuderna2024}. It may be crucial to use this information to realize the full potential of gLMs as foundation models.

In view of our results, we expect that the capabilities of gLMs will expand as larger whole-genome alignments are made available. We view this work as one motivation for doing so. For example, one way that PhyloGPN could have obtained stronger results on coding variant interpretation is by training on a whole-genome alignment that includes more distantly related species. Indeed, one of the differences between PhyloGPN and GPN-MSA is that the latter uses an alignment that includes more diverse vertebrate species but fewer species overall. This is one explanation for why PhyloGPN comes up short of GPN-MSA at predicting coding variant pathogenicity. We did not attempt to make use of this vertebrate alignment due to the technical challenge of harmonizing it with our mammalian alignment.

We note that our framework can, in principle, accommodate models of molecular evolution that are more comprehensive than the F81 model, such as the General Time-Reversible model \cite{tavare1986gtr}. This could be another way to improve the capabilities of gLMs.
Finally, it could be worthwhile to make use of gene trees to control for region-specific phylogenetic correlations.

\begin{credits}
\subsubsection{\ackname} We are grateful to Sebasti\'an Prillo and Junhao Xiong for providing helpful feedback on our methods and manuscript. This research is supported in part by an NIH grant R35-GM134922.

\subsubsection{Code and model availability.}
The code to reproduce the results of our log-likelihood-based evaluation is available at \url{https://github.com/songlab-cal/gpn}. The model is available at \url{https://huggingface.co/songlab/phylogpn}.

\end{credits}

\beginsupplement
\appendix
\section{Appendix}

\subsection{Mathematical Results}
\label{app:math}

\begin{proposition}
\label{prop:sigmoid_bound}
The inequality $1 - e^{-e^x} > \frac{e^x}{1 + e^x}$ holds for all $x \in \R.$
\label{prop:transition_prob_lower_bound}
\end{proposition}

\begin{proof}
Let $u = e^x.$ Since $u > 0$, the exponential bound $e^u \geq 1 + u$ is strict. It follows that $e^{-u} < (1 + u)^{-1}$ and that $1 - e^{-u} > \frac{u}{1 + u}$, which proves the result.
\end{proof}

\subsection{Computation of Log Likelihood Ratios}
\label{app:vep_evaluation_details}

Since PhyloGPN is a CNN, it can be applied to an input sequence as a sliding window, such that each likelihood is computed from exactly 481 bp of context. On the other hand, our baselines require separate forward passes for each position to make use of their entire receptive field. To expedite their evaluation, we partitioned the human reference genome into disjoint 3 kb segments and obtained likelihood values from each model for each of these segments by applying the model to the largest possible input sequence with the segment at the center. As a result, Nucleotide Transformer (NT) used at least 9,283 bp of context around each position instead of the maximum-possible 12,281 bp to compute its likelihood values; Caduceus used at least 128,071 bp instead of 131,071 bp; and HyenaDNA used at least 314,001 bp instead of 320,001 bp.

We used \texttt{nucleotide-transformer-v2-500m-multi-species}, available on HuggingFace,
to evaluate NT. We also evaluated the \texttt{nucleotide-transformer-} \texttt{2.5b-multi-species} model, since this one yielded better results on the Disease VEP task from BEND. However, we found that this model performed slightly worse on likelihood-based tasks. NT tokenizes its inputs by mapping each disjoint 6-mer to one of 4,104 tokens. To obtain the LLR of a substitution at a given position of GRCh38, we first apply NT to a segment containing the position and generate likelihoods for each position of the tokenized segment. We then obtain the LLR of the token corresponding to the alternate allele over that of the reference. We do the same for the reverse complement of the segment, obtaining a LLR for the complementary substitution. We use the log geometric mean of these two ratios---equivalently, the arithmetic mean of the LLRs---as the final score of the substitution. We observed that this did not make the performance of the model any worse than only using the first LLR as the score.

To evaluate HyenaDNA, we used \texttt{hyenadna-medium-160k-seqlen-hf}, which was the largest model that was feasible for us to evaluate. Because HyenaDNA is a causal language model, it computes likelihoods for given positions solely from the sequence context preceding the position. To put it on the same footing as our other baselines, we also apply it to the reverse complement of the nucleotides following the position, and average the resulting LLRs as we did with NT.

To evaluate Caduceus, we used the \texttt{caduceus-ps\_seqlen-131k\_d\_model-} \texttt{256\_n\_layer-16} model. We did not transform its outputs, as it already uses bidirectional context and is reverse-complement equivariant.

Lastly, we evaluated GPN-MSA using pre-computed LLR values. These data are available at https://huggingface.co/datasets/songlab/gpn-msa-hg38-scores.

\subsection{Evaluation of GPN-MSA on BEND}
\label{subsec:gpn-msa-bend}

To obtain embeddings for the BEND tasks, we first used the genome coordinates provided for each sequence to retrieve the corresponding alignments from the multiple sequence alignment (MSA) generated by GPN‑MSA. Because the sequences in the BEND tasks exceed GPN‑MSA’s context size of 128 bp, we divided each alignment into non-overlapping 128 bp windows, computed embeddings for each window individually, and then concatenated these window embeddings to reconstruct the full-sequence embedding. Note that, since the MSA was constructed using the hg39 human reference genome, we could only evaluate GPN‑MSA on all BEND tasks except Chromatin Accessibility, which is based on hg19 coordinates. After obtaining the GPN‑MSA embeddings, we followed the default BEND configurations for training and evaluating both the classification tasks and the variant effect prediction tasks.

\subsection{Supplementary Figure and Tables}

\begin{figure}[h!]
\centering
\includegraphics[trim = 0 4mm 0 0, clip, width=0.6\textwidth]{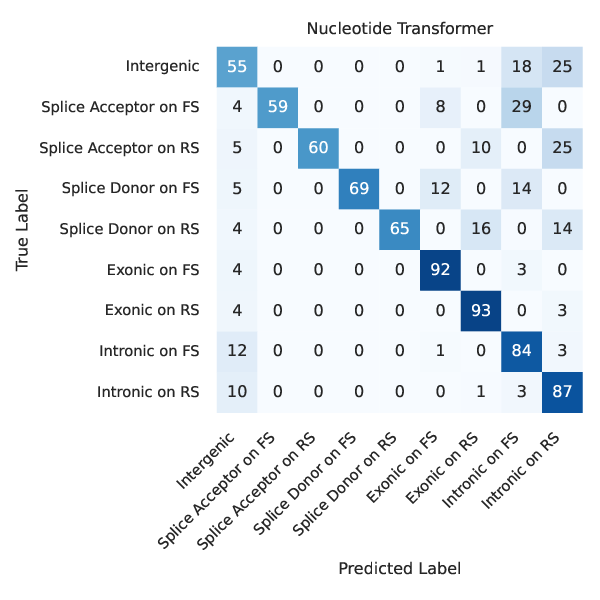}
\caption{Nucleotide Transformer's confusion matrix for the Gene Finding task.}
\label{fig:gene_finding_extended}
\end{figure}

\newpage

\begin{table}[t]
\centering
\caption{Results on the task of predicting variant rankings from DMS experiments. Values are Spearman correlations. For each experiment, the best result among all gLMs except GPN-MSA is typeset in boldface. If the result for GPN-MSA is best overall, it is also typeset in boldface. Asterisks denote data from the study Giacomelli et al. 2018.}
\label{tab:dms_results}
{\scriptsize
\begin{tabularx}{\textwidth}{lCCCC!{\vrule width 0.5pt}C}\toprule
 &  & Nucleotide & & \\
ID & PhyloGPN & Transformer & HyenaDNA & Caduceus & GPN-MSA \\
\midrule
A4 & 0.42 & \textbf{0.50} & $-$0.08\phantom{$-$} & $-$0.04\phantom{$-$} & 0.37 \\
ADRB2 & \textbf{0.35} & 0.30 & 0.05 & 0.04 & \textbf{0.43} \\
BRCA1 & \textbf{0.30} &  $-$0.09\phantom{$-$} &  $-$0.18\phantom{$-$} & $-$0.10\phantom{$-$} & \textbf{0.36} \\
CALM1 & \textbf{0.11} & 0.00 & $-$0.01\phantom{$-$} & 0.00 & \textbf{0.15} \\
CYP2C9 (Abundance) & \textbf{0.30} & 0.04 & $-$0.05\phantom{$-$} & 0.03 & \textbf{0.49} \\
CYP2C9 (Activity) & \textbf{0.24} & 0.04 & $-$0.03\phantom{$-$} & 0.01 & \textbf{0.50} \\
DLG4 & \textbf{0.47} & 0.24 & $-$0.09\phantom{$-$} & 0.29 & \textbf{0.51} \\
GRB2 & \textbf{0.30} & 0.24 & $-$0.06\phantom{$-$} & 0.10 & \textbf{0.31} \\
KCNH2 & \textbf{0.37} & $-$0.07\phantom{$-$} & 0.19 & $-$0.02\phantom{$-$} & 0.29 \\
MAPK1 & \textbf{0.11} & 0.08 & $-$0.26\phantom{$-$} & $-$0.10\phantom{$-$} & \textbf{0.15} \\
MSH2 & \textbf{0.18} & 0.06 & 0.03 & $-$0.03\phantom{$-$} & \textbf{0.27} \\
NUD15 & \textbf{0.16} & 0.02 & 0.09 & $-$0.04\phantom{$-$} & \textbf{0.53} \\
TP53 (Null, Etoposide)* & \textbf{0.34} & 0.02 & 0.07 & $-$0.10\phantom{$-$} & \textbf{0.39} \\
TP53 (Null, Nutlin)* & \textbf{0.34} & 0.02 & 0.08 & $-$0.08\phantom{$-$} & 0.35 \\
TP53 (WT, Nutlin)* & \textbf{0.50} & $-$0.04\phantom{$-$} & $-$0.02\phantom{$-$} & $-$0.11\phantom{$-$} & 0.45 \\
TP53 (Kotler et al. 2018) & \textbf{0.29} & 0.01 & 0.02 & $-$0.06\phantom{$-$} & \textbf{0.50} \\
PTEN (Matreyek et al. 2021) & \textbf{0.24} & 0.08 & 0.05 & 0.00 & \textbf{0.26} \\
PTEN (Mighell et al. 2018) & \textbf{0.15} & 0.07 & 0.02 & $-$0.04\phantom{$-$} & \textbf{0.23} \\
SLC6A4 & \textbf{0.31} & 0.24 & $-$0.06\phantom{$-$} & $-$0.00\phantom{$-$} & \textbf{0.37} \\
SCN5A & \textbf{0.11} & 0.10 & $-$0.15\phantom{$-$} & 0.02 & 0.18 \\
SRC & \textbf{0.15} & 0.11 & $-$0.02\phantom{$-$} & $-$0.01\phantom{$-$} & 0.16 \\
SUMO1 & \textbf{0.15} & $-$0.11\phantom{$-$} & 0.01 & $-$0.08\phantom{$-$} & \textbf{0.17} \\
SNCA & \textbf{0.21} & 0.06 & 0.12 & 0.00 & \textbf{0.29} \\
TARDBP & \textbf{0.36} & 0.11 & $-$0.03\phantom{$-$} & $-$0.03\phantom{$-$} & \textbf{0.48} \\
TPK1 & \textbf{0.18} & 0.02 & $-$0.05\phantom{$-$} & 0.02 & \textbf{0.22} \\
TPMT & \textbf{0.32} & 0.05 & 0.07 & $-$0.04\phantom{$-$} & \textbf{0.46} \\
TPOR & \textbf{0.43} & 0.15 & 0.08 & $-$0.03\phantom{$-$} & 0.36 \\
UBE2I & \textbf{0.18} & $-$0.10\phantom{$-$} & $-$0.05\phantom{$-$} & $-$0.11\phantom{$-$} & \textbf{0.20} \\
VKOR1 (Abundance) & \textbf{0.24} & 0.02 & 0.22 & $-$0.02\phantom{$-$} & \textbf{0.39} \\
VKOR1 (Activity) & \textbf{0.24} & 0.08 & 0.15 & 0.12 & \textbf{0.38} \\
YAP1 & \textbf{0.09} & 0.06 & 0.02 & 0.04 & \textbf{0.31} \\
\bottomrule
\end{tabularx}
}
\end{table}

\begin{table}[t]
\centering
\caption{Ablation study results on the task of predicting ClinVar classifications stratified by variant class. “Subset 1” consists of all positions in the training data of PhyloGPN that
are in odd-numbered chromosomes or the X chromosome. “Subset 2” consists of all the other positions in
the training data. The baseline model was trained on all the positions in the training data for one epoch. The ablated model was trained on Subset 1 for two epochs. Models are evaluated on Subset 2. Each entry is an AUROC. The best result for each category is typeset in boldface. }
\label{tab:ablation_results}
{\scriptsize
\begin{tabularx}{\textwidth}{lCC}\toprule
Category & Baseline Model & Ablated Model \\
\midrule
3' UTR & 0.84 & \textbf{0.87} \\
5' UTR & 0.81 & \textbf{0.83} \\
All & 0.85 & \textbf{0.87} \\
Downstream of Transcript & \textbf{0.76} & 0.70 \\
Intronic & 0.81 & \textbf{0.84} \\
Missense & \textbf{0.72} & \textbf{0.72} \\
Noncoding & 0.82 & \textbf{0.85} \\
Splice Acceptor & 0.69 & \textbf{0.71} \\
Splice Donor & 0.70 & \textbf{0.71} \\
Start Codon & \textbf{0.49} & \textbf{0.49} \\
Stop Lost & 0.65 & \textbf{0.67} \\
Synonymous & 0.64 & \textbf{0.71} \\
Upstream of Transcript & 0.68 & \textbf{0.75} \\
\bottomrule
\end{tabularx}
}
\end{table}


\clearpage

\bibliographystyle{splncs04}

\end{document}